\documentclass[prl,aps,showpacs,twocolumn,floatfix]{revtex4}
\usepackage{epsfig} \usepackage{graphics} \usepackage{bm}
\usepackage{amssymb}
\usepackage{graphicx}
\begin{document}

\preprint{Natala-PRL}

\title{Strongly Forbidden Thermodynamic
Oscillations in Quasi-One-Dimensional Conductors}

\author{A.G. Lebed}

\affiliation{Department of Physics, University of Arizona, 1118 E.
4-th Street, Tucson, AZ 85721, USA}

\begin{abstract}
We theoretically show that strongly forbidden oscillations of a
specific heat have to exist in metallic phases of some
quasi-one-dimensional (Q1D) conductors. They appear due to
electron-electron interactions under condition of the magnetic
breakdown phenomenon between the so-called open interference
electron orbits. We argue that such forbidden thermodynamic
oscillations can exist in Q1D conductors (TMTSF)$_2$ClO$_4$ and
(Per)$_2$Au(mnt)$_2$, where TMTSF stands for
tetramethyltetraselenafulvalene, Per is polycyclic aromatic
hydrocarbon and mnt is mononitrotoluene,
 and suggest to discover them.
\end{abstract}


\maketitle

It is well known that, in layered quasi-one-dimensional (Q1D)
conductors, closed quasi-particle orbits do not exist in a
magnetic field. This prevents the appearance of quantum effects
due to the so-called Landau quantization [1] of electron energy
levels in the field. Nevertheless, in magnetic fields in Q1D
conductors, there are some other quantum effects - the Bragg
reflections of electrons from the Brillouin zones boundaries
[2-5]. They cause the existence in (TMTSF)$_2$- and (ET)$_2$-based
Q1D conductors, where TMTSF stands for
tetramethyltetraselenafulvalene and ET stands for the so-called
ethyl group, of such quantum phases as the
Field-Induced-Spin-Density-Wave (FISDW) ones, exhibiting 3D
Quantum Hall effect [5]. Moreover, the above mentioned conductors
demonstrate some exotic angular conductivity oscillations of
quantum interference origin such as Lebed's Magic Angles (LMA),
Third Angular Effect (TAE), and Lee-Naughton-Lebed's (LNL) angles
in their metallic phases (for a review, see Refs. [5,6]).
According to the most of current theories, some LMA, TAE, and LNL
angular oscillations can be explained by the Bragg reflections of
non-interacting electrons within the Fermi liquid (FL) approach
[1,5].

Meanwhile, as was shown theoretically [7-10], interactions of Q1D
electrons can result in weak [7,8] and the strongest [9,10]
deviations from the FL theory in magnetic fields. Indeed, as shown
in Ref.[7], some novel oscillations appear in kinetic properties,
like conductivity, whereas in Ref.[9] it was shown by Yakovenko
that the similar to [7] angular and magnetic oscillations could
appear even in thermodynamic properties such as magnetic moment of
electrons moving along open orbits. Note that the last statement
strongly contradicts the FL theory [1]. Due to small amplitudes of
the predicted non-FL oscillations, the non-FL effects [9,10] have
not been observed yet in Q1D metals. The next important step in
the theory [11] was the consideration of the weak deviations from
the FL results for open electron trajectories in magnetic fields
under the condition of magnetic breakdown between open electron
orbits (i.e., due to the so-called Stark effect [12-17]). It was
shown [11] that electron-electron scattering time oscillations
were much increased in their magnitudes and such oscillations were
probably experimentally observed in resistivity measurements in
(TMTSF)$_2$ClO$_4$ [18].

The goal of our Letter is to study theoretically influence of the
Stark variant of the magnetic breakdown on the most principle
violations of the FL theory - the appearance of the forbidden
thermodynamical oscillations. In particular, we show that
electron-electron interactions cause the existence of the
forbidden specific heat oscillations in metallic phases of some
Q1D conductors even in the absence of closed quasi-particles
orbits in a magnetic field. The amplitudes of such oscillations
are highly enlarged, if we compare them to the oscillations [9].
Physical origin of the oscillations of specific heat is an
oscillatory nature of electron spectrum under the condition of
magnetic breakdown [15-17], where the corrections to specific heat
can be considered as strong fluctuations which, as shown by us
below, exist far above the FISDW Peierls phase transition.
Therefore, we hope that they can be observed in a metallic phase
of Q1D conductor (TMTSF)$_2$ClO$_4$ in experiments similar to the
more recent experiment [19]. Note that there is a principle
difference between our current calculations and the results of
Refs. [15-17]. In this Letter, we calculate corrections to
specific heat in a metallic phase, whereas all previous
calculations were performed in FISDW phase. From mathematical
point of view, this means that our current calculations involve a
product of four Green's functions, in contrast to the case
[15-17], where only products of two Green's functions were
considered. From physical point of view, we calculate forbidden in
the FL theory oscillations, in contrast to Refs. [15-17], where
allowed in the FL theory oscillations were considered. Another
candidate for the predicted by us non FL effects is Q1D conductor
(Per)$_2$Au(mnt)$_2$, where Per is polycyclic aromatic hydrocarbon
and mnt is mononitrotoluene, which also exhibits the Stark effect
[20].

\begin{figure}[t]
\centering
\includegraphics[width=0.5\textwidth]{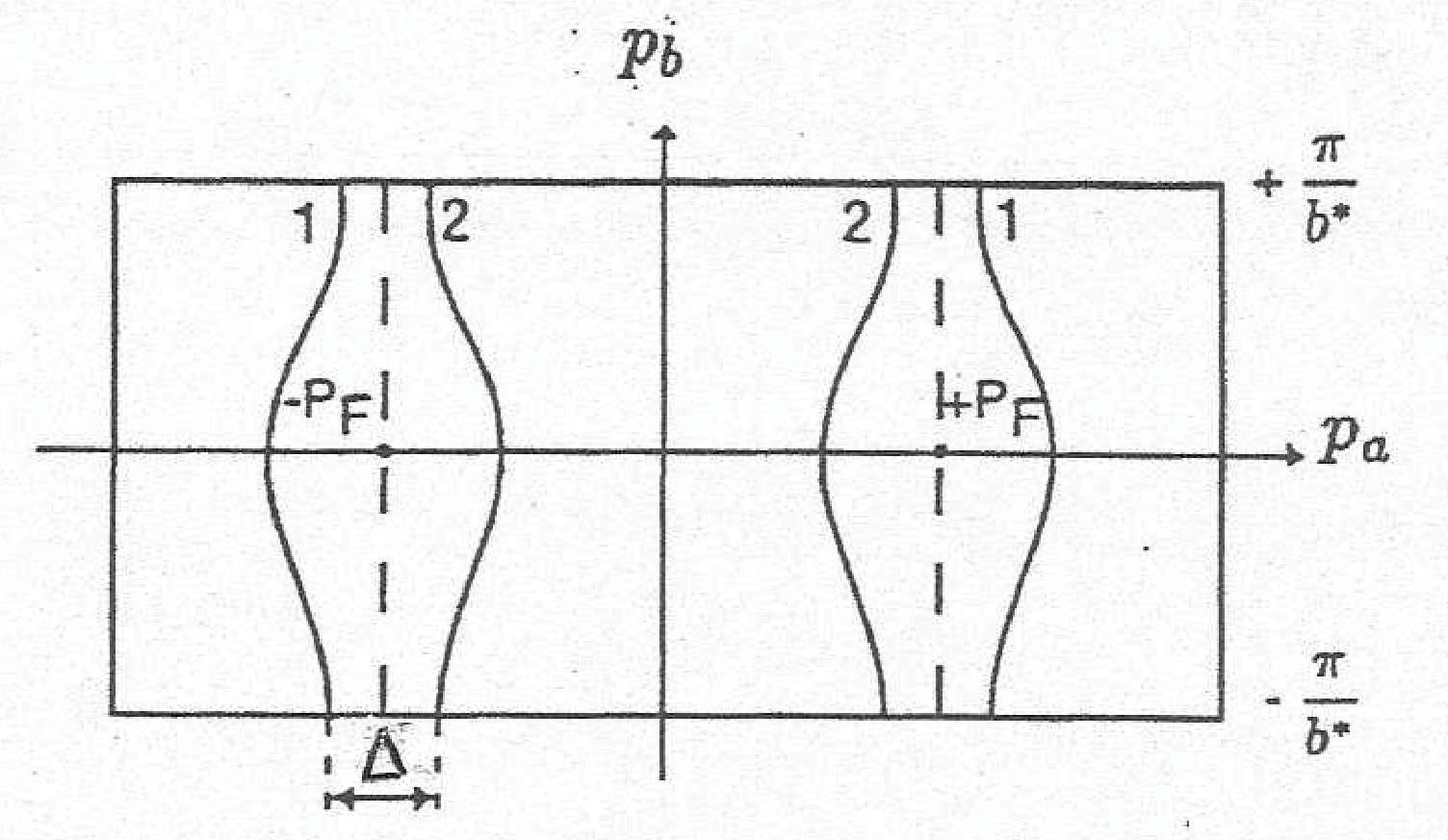}
\caption{Quasi-one-dimensional Fermi surface of the organic
conductor (TMTSF)$_2$ClO$_4$ in the presence of anion ordering
gap, $\Delta$ [see Eq.(4)].}
\end{figure}

Note that, in the absence of the so-called anion gap [14], Q1D
electron spectrum of the (TMTSF)$_2$ClO$_4$ conductor can be
written as [5]
\begin{equation}
\epsilon^{\pm}({\bf p})= \pm v_F (p_x \mp p_F) + 2 t_b \cos(p_y
b^*)] + 2 t_c \cos(p_z c^*),
\end{equation}
where $v_F p_F \gg t_b \gg t_c$; $b^*$ and $c^*$ are crystalline
lattice parameters. The anion gap, $\Delta(y)= \Delta \cos(\pi
y/b^*)$, leads to a doubling of the lattice periodicity along
${\bf y}$ axis and the electron wave functions obey the following
equations:
\begin{eqnarray}
[\pm v_F (p_x \mp p_F) + 2t_b \cos(p_y
b^*)]\psi^{\pm}_{\epsilon}(p_y)
\nonumber\\
 + \Delta \ \psi^{\pm}_{\epsilon}(p_y +\pi/b^*)=
 \epsilon \psi^{\pm}_{\epsilon}(p_y),
\end{eqnarray}
\begin{eqnarray}
[\pm v_F (p_x \mp p_F) - 2t_b \cos(p_y
b^*)]\psi^{\pm}_{\epsilon}(p_y+\pi/b^*)
\nonumber\\
 + \Delta \ \psi^{\pm}_{\epsilon}(p_y)=\epsilon \psi^{\pm}_{\epsilon}(p_y+\pi/b^*).
\end{eqnarray}
As a result, in the presence of the anion gap there exist the
following four sheets of the Q1D Fermi surface (see Fig.1):
\begin{eqnarray}
&&\epsilon^{\pm}_{n}({\bf p})= \pm v_F (p_x \mp p_F)
\nonumber\\
&&+(-1)^n \sqrt{[2 t_b \cos(p_y b^*)]^2 + \Delta^2}, \ \ \, n=1,2,
\end{eqnarray}
which correspond to the real experimental situation in the
(TMTSF)$_2$ClO$_4$ at ambient pressure. [Note that here we
disregard the term $2t_c \cos (p_z c^*)$, but account for it at
the end of the Letter in our final equations.]

Let us now perform the so-called Peierls substitution [1,5,13],
\begin{equation}
p_x \mp p_F \rightarrow - i \frac{d}{dx}, \ \ \ p_y \rightarrow
p_y - \frac{e}{c} A_y,
\end{equation}
 in
Eqs.(2) and (3) in the external magnetic field perpendicular to
conducting plane $({\bf a},{\bf b^*})$ :
\begin{equation}
{\bf H}=(0,0,H), \ \ \ {\bf A}=(0,Hx,0).
\end{equation}
[Note that in this Letter we use system units where the Planck
constant $\hbar$=1]. In this case, Eqs.(2) and (3) can be
rewritten as

\begin{eqnarray}
\biggl[ \mp i v_F \frac{d}{dx} + 2t_b \cos \biggl( p_y b^*
-\frac{\omega_c x}{v_F} \biggl)
\biggl]\psi^{\pm}_{\epsilon}(p_y,x)
\nonumber\\
 + \Delta \ \psi^{\pm}_{\epsilon}(p_y +\pi/b^*,x)=
 \epsilon \psi^{\pm}_{\epsilon}(p_y,x),
\end{eqnarray}

\begin{eqnarray}
\biggl[ \mp i v_F \frac{d}{dx} - 2t_b \cos \biggl( p_y b^*
-\frac{\omega_c x}{v_F} \biggl)
\biggl]\psi^{\pm}_{\epsilon}(p_y+\pi/b^*,x)
\nonumber\\
 + \Delta \ \psi^{\pm}_{\epsilon}(p_y,x)=
 \epsilon \psi^{\pm}_{\epsilon}(p_y+\pi/b^*,x),
\end{eqnarray}
where $\omega_c=eHv_Fb^*/c$ is the so-called cyclotron frequency
of electron motion along open electron trajectories in the
Brillouin zone [2,5].

Note that magnetic breakdown problem of Eqs. (7) and (8) was
carefully studied in Ref.[17] where the magnetic breakdown field
was calculated,
\begin{equation}
H_{MB} = \frac{\pi c \Delta^2}{2ev_F t_b b^*}.
\end{equation}
Below, we consider the case of very high magnetic fields
[11,15,16],
\begin{equation}
H \geq H_{MB} ,
\end{equation}
where we can use the perturbation approach with respect to the
anion ordered gap for solutions of Eqs.(7) and (8). In this case,
in the first approximation wave functions are symmetric (11) and
anti-symmetric (12) combinations of two solutions of Eqs.(7) and
(8) with $\Delta=0$ with opposite energy shifts due to $\Delta
\neq 0$. As a result, we obtain the following two component
vector[11]:

\begin{eqnarray}
&[\psi^{\pm}_1(p_y,x),\psi^{\pm}_1(p_y+\pi/b^*,x)]=\frac{\exp[\pm
i(\epsilon-\Delta^*)x/v_F]}{\sqrt{2}}
\nonumber\\
&\biggl\{\exp \biggl[\pm \frac{i \lambda}{2} \sin
\biggl(p_yb^*-\frac{\omega_c x}{v_F} \bigg) \bigg],\exp
\biggl[\mp\frac{i \lambda}{2} \sin \biggl(p_yb^*-\frac{\omega_c
x}{v_F} \bigg) \bigg] \biggl\}
\end{eqnarray}
and
\begin{eqnarray}
&[\psi^{\pm}_2(p_y,x),\psi^{\pm}_2(p_y+\pi/b^*,x)]=\frac{\exp[\pm
i(\epsilon+\Delta^*)x/v_F]}{\sqrt{2}}
\nonumber\\
&\biggl\{\exp \biggl[\pm \frac{i \lambda}{2} \sin
\biggl(p_yb^*-\frac{\omega_c x}{v_F} \bigg) \bigg],-\exp
\biggl[\mp\frac{i \lambda}{2} \sin \biggl(p_yb^*-\frac{\omega_c
x}{v_F} \bigg) \bigg] \biggl\},
\end{eqnarray}
where
\begin{equation}
\lambda = \frac{4t_b}{\omega_c}, \ \ \ \ \omega_c = ev_FH b^*/c.
\end{equation}
Note that the symmetric (11) and anti-symmetric (12) wave
functions have different energies [11,15,16],
\begin{equation}
\epsilon^{\pm}_1({\bf p})= \epsilon  - \Delta^*, \ \ \epsilon =
\pm v_F(p_x \mp p_F),
\end{equation}
\begin{equation}
\epsilon^{\pm}_2({\bf p}) = \epsilon + \Delta^*, \ \ \epsilon =
\pm v_F(p_x \mp p_F),
\end{equation}
with the difference in energies, 2$\Delta^*$, being an oscillating
function of an inverse magnetic field :
\begin{equation}
\Delta^* = J_0(\lambda) \Delta \simeq \Delta \
\sqrt{\frac{\omega_b}{2 \pi t_b}}\cos \biggl( \frac{4t_b c}{ev_F H
b^*} \biggl) ,
\end{equation}
where $J_0(...)$ is the zeroth order Bessel function. It is
important that the period of the oscillations of $(\Delta^*)^2$
(16) is equal to
\begin{equation}
\delta \biggl( \frac{1}{H} \biggl) = \frac{\pi ev_F b^*}{4t_b c} ,
\end{equation}
and, as shown below, the specific heat correction due to
electron-electron interactions in a metallic phase oscillates
exactly with this period.
\begin{figure}[t]
 \centering
\includegraphics[width=0.35\textwidth]{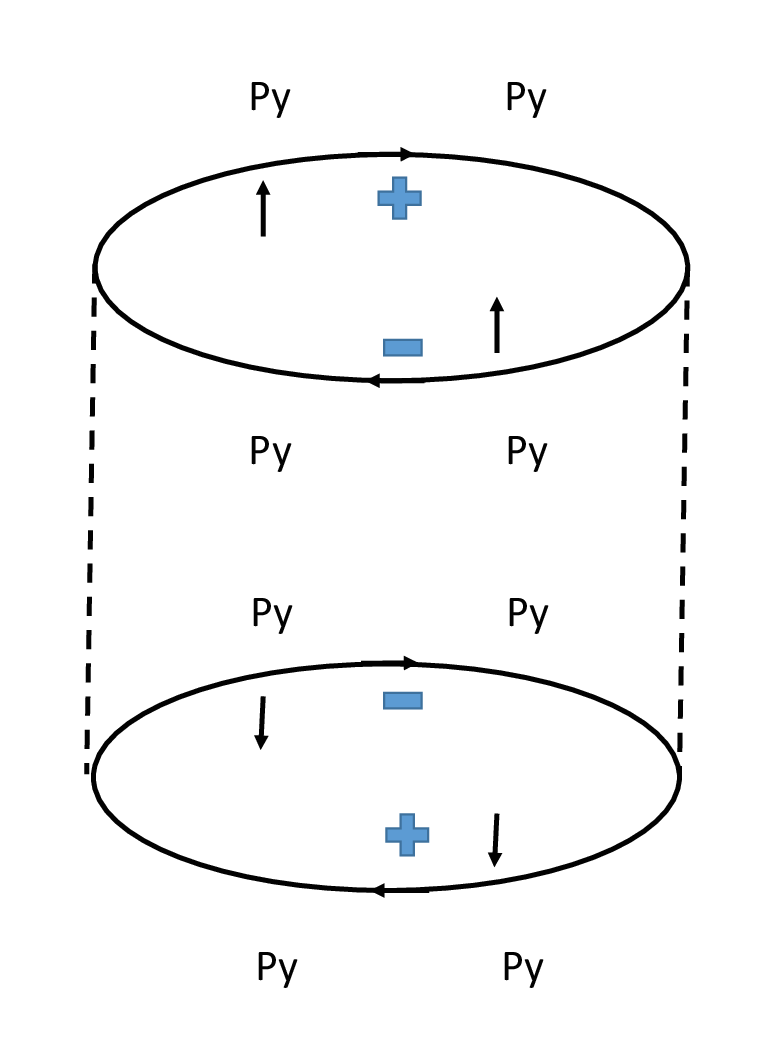}
\caption{Diagram 1: Feynman diagram of interacting
quasi-one-dimensional electrons in the presence of the anion
ordering gap [see Eq.(4)]. where electrons penetrate through the
gap in strong magnetic fields. Electron Green functions are shown
by solid lines, where the electron-electron interactions are shown
by broken lines.}
\end{figure}
\begin{figure}[t]
\centering
\includegraphics[width=0.35\textwidth]{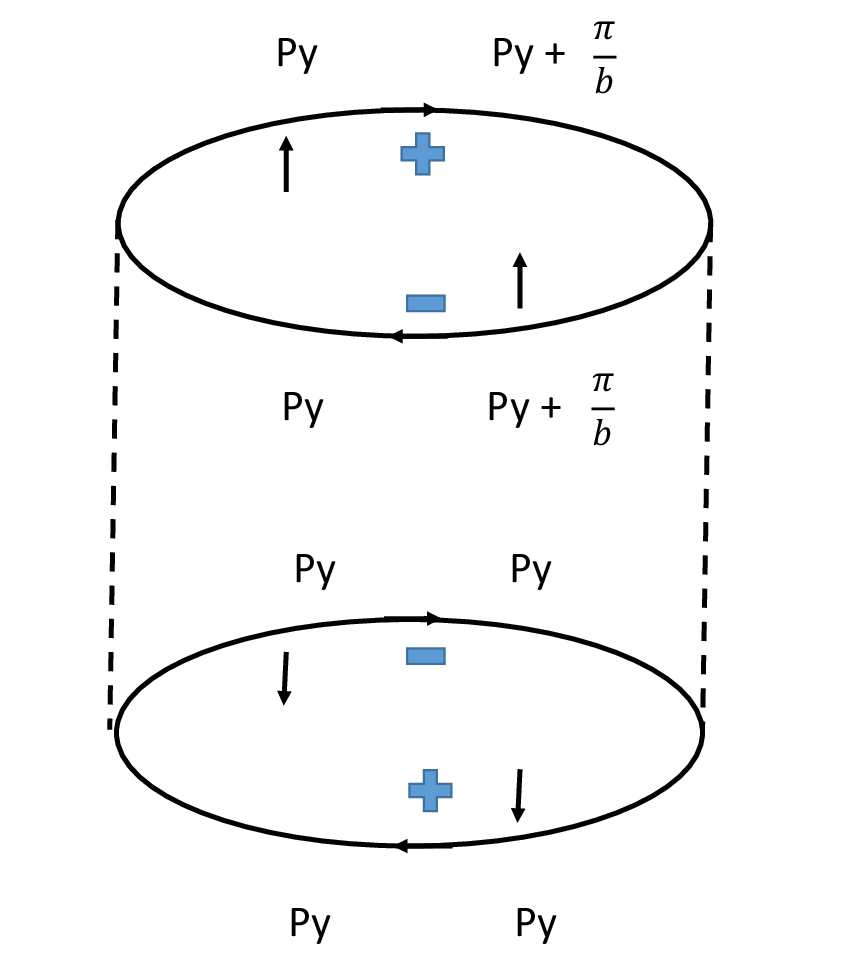}
\caption{Diagram 2: Another Feynman diagram of interacting
quasi-one-dimensional electrons in the presence of the anion
ordering.}
\end{figure}
To calculate corrections to the free energy of a metallic phase
due to electron-electron interactions, we make use of the method
of the Matsubara Green functions [21]. Once wave functions and
energy spectrum are known [see Eqs.(11)-(15)], we can derive the
Matsubara Green functions using the following standard procedure:
\begin{eqnarray}
G_{\pm}(i \omega_n; x,x'; p_y,p_y)= \sum_{j=1,2}
\sum_{\epsilon^{\pm}_j} \nonumber\\
\frac{[\psi^{\pm}_j(\epsilon^{\pm}_j
;x,p_y)]^*\psi^{\pm}_j(\epsilon^{\pm}_j; x,p_y)}{i \omega_n -
\epsilon^{\pm}_j}
\end{eqnarray}
and
\begin{eqnarray}
G_{\pm}(i \omega_n; x,x'; p_y,p_y+\frac{\pi}{b^*})=
\sum_{j=1,2} \sum_{\epsilon^{\pm}_j} \nonumber\\
\frac{[\psi^{\pm}_j(\epsilon^{\pm}_j;
x,p_y)]^*\psi^{\pm}_j(\epsilon^{\pm}_j; x,p_y+\frac{\pi}{b^*})}{i
\omega_n - \epsilon^{\pm}_j},
\end{eqnarray}
where $\omega_n = 2 \pi T (n+1/2)$ is the Matsubara frequency for
fermions [21]. Note that below we consider the case of high
magnetic fields (10), therefore, we use the approximation of
Ref.[15,16] to calculate the Green's functions. This approximation
considers the magnetic breakdown phenomenon as a perturbation
which splits the electron wave spectrum into two branches with
energies (14),(15). As a result, we obtain [15]:
\begin{figure}[t]
\centering
\includegraphics[width=0.35\textwidth]{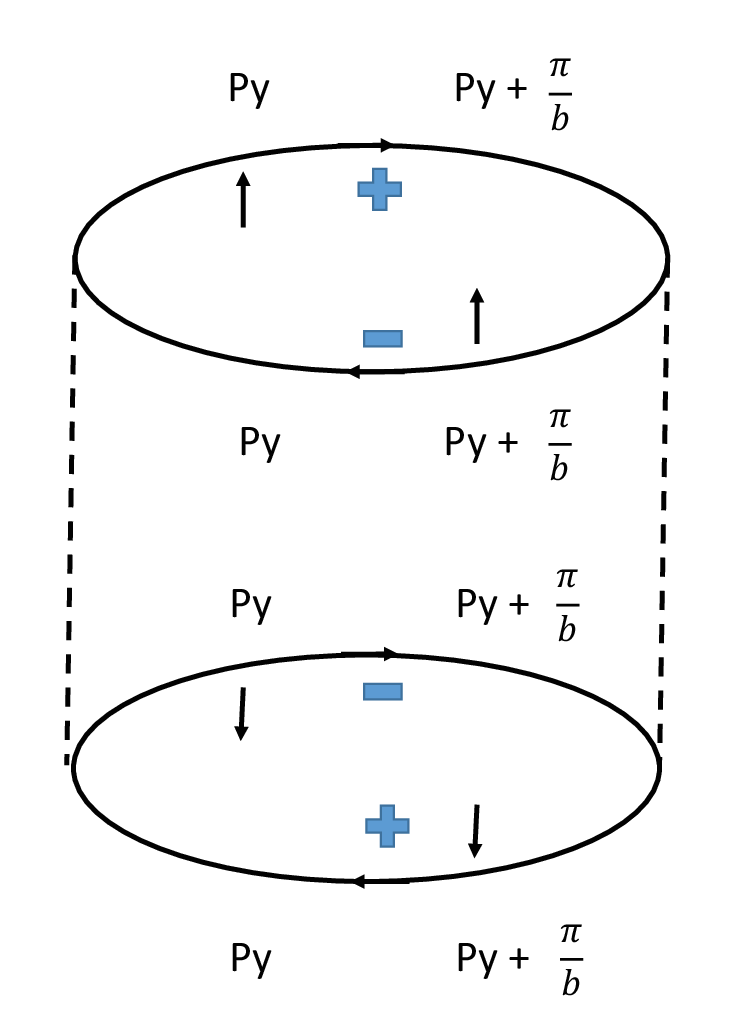}
\caption{Diagram 3: One more Feynman diagram of interacting
quasi-one-dimensional electrons in the presence of the anion
ordering gap [see Eq.(4)].}
\end{figure}

\begin{eqnarray}
&G_{+} \biggl(i \omega_n; p_x,p_x+\frac{\omega_c l}{v_F};p_y,p_y
\biggl)=
\nonumber\\
&\sum^{+\infty}_{m=-\infty}
\frac{J_m(\lambda)J_{m+l}(\lambda)\exp(ip_ylb)(i \omega_n -p_xv_F
-\omega_c m)}{(i \omega_n -p_x v_F-\omega_c m)^2-J^2_0(\lambda)
\Delta^2} ,
\end{eqnarray}
\begin{eqnarray}
&G_{+} \biggl(i \omega_n; p_x,p_x+\frac{\omega_c
l}{v_F};p_y,p_y+\frac{\pi}{b} \biggl)=
\nonumber\\
&\sum^{+\infty}_{m=-\infty}
\frac{J_m(\lambda)J_{m+l}(\lambda)\exp(ip_ylb)J_m(\lambda)\Delta}{(i
\omega_n -p_x v_F-\omega_c m)^2-J^2_0(\lambda) \Delta^2} .
\end{eqnarray}

\begin{figure}[t]
\centering
\includegraphics[width=0.35\textwidth]{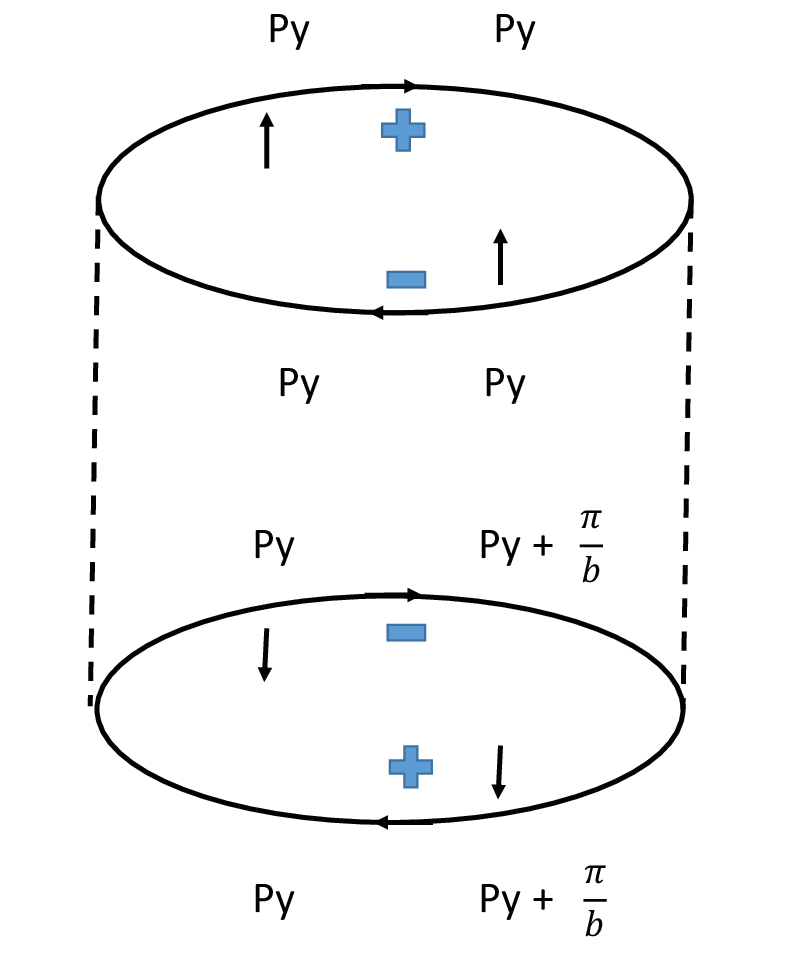}
\caption{Diagram 4: the last Feynman diagram of interacting
quasi-one-dimensional electrons in the presence of the anion
ordering gap [see Eq.(4).}
\end{figure}
\begin{figure}[t]
\centering
\includegraphics[width=0.50\textwidth]{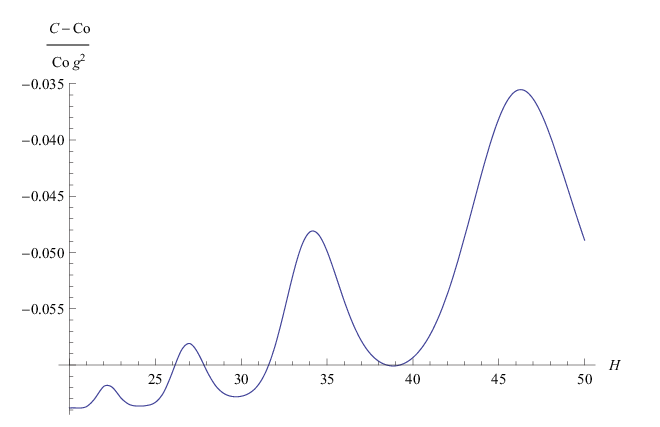}
\caption{Normalized correction to a specific heat of the
(TMTSF)$_2$ClO$_4$ conductor numerically calculated from Eq.(23).}
\end{figure}

Four possible contributions to the electron free energy due to
electron-electron interactions in the presence of the anion gap
are shown in Fig.2, Fig.3, Fig4, and Fig.5. Calculating all of
such diagrams, which do not contain spin-splitting of the energy
in a magnetic field, we obtain the following formula for the
contribution to the free energy per one electron due to the
electron-electron interactions:

\begin{eqnarray}
\delta F(H) = -\frac{g^2 \pi^3 T^3}{p_F v^2_F} \int^{\infty}_{0}
dx \frac{\cosh(2 \pi T x/v_F)}{\sinh^3(2 \pi T x/v_F)} \cos^4
\biggl(\frac{\Delta^* x}{v_F} \biggl)
\nonumber\\
\times \biggl<J^2_0 \biggl[2 \lambda \sin \biggl(\frac{\omega_c
x}{v_F} \bigg) \sin(p)\biggl] \biggl>_{p}
 \biggl<J^2_0 \biggl[\biggl(\frac{4t_c x}{v_F} \bigg)
\sin(k)\biggl] \biggl>_{k} ,
\end{eqnarray}
where $g$ is a dimensionless constant of the electron-electron
interactions, the Boltzmann constant is $k_B=1$. Note that the
brackets $<...>_p$ and $<...>_k$ in Eq.(22) stand for averaging
procedure over $p$ and $k$, respectively.

 We point out that Eq.(22)
diverges at x=0. Nevertheless it is possible to show that the
corresponding correction to a specific heat is a convergent
function and is equal to
\begin{eqnarray}
C-C_0 = -\frac{3}{4} g^2 C_0 \int^{\infty}_{0} dx \biggl(
\frac{x^2}{\sinh^2(x)} \biggl)''' \cos^4 \biggl(\frac{2 \Delta^*
x}{4\pi T} \biggl)
\nonumber\\
\times \biggl<J^2_0 \biggl[2 \lambda \sin \biggl(\frac{\omega_c
x}{4\pi T} \bigg) \sin(p)\biggl] \biggl>_{p}
 \biggl<J^2_0 \biggl[\biggl(\frac{2t_c x}{\pi T} \bigg)
\sin(k)\biggl] \biggl>_{k},
\end{eqnarray}
where $C_0$ is specific heat of non-interacting electrons,
\begin{equation}
C_0=\frac{\pi^2}{3} \frac{T}{p_Fv_F}.
\end{equation}

Let us calculate the correction to specific heat (23) numerically.
For this purpose we use the following vales of the parameters:
$t_c=2.5 \ K$ [22], $\Delta = 40 \ K$ [11], and $\omega_c(H)/H = 2
\ K/T$ [5]. As a result, we obtain the following oscillatory
behavior between $20 \ T$ and $50 \ T$ (see Fig.4), where the
magnitude of the oscillations is quickly rising function of a
magnetic field and can achieve the value $\delta C/C_0 \simeq
10^{-2}$. We suggest to perform the corresponding experiments in
Q1D organic conductor (TMTSF)$_2$ClO$_4$, whose band parameters
have been used for the calculations, and in Q1D organic conductor
(Per)$_2$Au(mnt)$_2$, whose band parameters are not such well
known. Here, we discuss the experimental conditions to be
fulfilled in (TMTSF)$_2$ClO$_4$ for the observation of the
forbidden specific heat oscillations. First of all, the
temperature has to be $T \geq 5K$ in order that the above
mentioned compound will be in the metallic phase. Secondly,
magnetic fields have to be of the order of $H \simeq 20-50 \ T$,
since we have made all calculations under the condition (10),
where $H_{MB} \simeq 10-15 \ T$ [1]. To the best of our knowledge
the forbidden oscillations of the specific heat due to magnetic
breakdown neither have been theoretically calculated nor have been
experimentally observed before.

The author is thankful to N.N. Bagmet (Lebed) for useful
discussions.

\end{document}